\documentclass[review]{elsarticle}

\usepackage{lineno}
\usepackage{mathtools}
\usepackage{amsmath}
\usepackage{color}
\usepackage{xcolor}
\usepackage{bm}
\usepackage{mathrsfs}
\usepackage{amssymb}
\usepackage{euscript}
\usepackage{algorithm}
\usepackage{algorithmic}
\usepackage{graphicx}
\usepackage{subcaption}
\usepackage{multirow}
\usepackage[Symbol]{upgreek}
\usepackage{booktabs}
\usepackage{makecell}
\usepackage[english]{babel}
\usepackage{diagbox}
\usepackage{gensymb}

\modulolinenumbers[1]

\journal{XX}

\begin{document}

\begin{frontmatter}

\title{Noise-robust blind reverberation time estimation using noise-aware time-frequency masking}
\author[mysecondaryddress,mythirdaddress]{Kaitong Zheng}
\author[mysecondaryddress,mythirdaddress]{Chengshi Zheng}
\author[mysecondaryddress,mythirdaddress]{Jinqiu Sang\corref{mycorrespondingauthor}}
\cortext[mycorrespondingauthor]{Corresponding author}
\ead{sangjinqiu@mail.ioa.ac.cn}
\author[mysecondaryddress,mythirdaddress]{Yulong Zhang}
\author[mysecondaryddress,mythirdaddress]{Xiaodong Li}

\address[mysecondaryddress]{Key Laboratory of Noise and Vibration Research, Institute of Acoustics, Chinese Academy of Sciences,
100190, Beijing, China}
\address[mythirdaddress]{University of Chinese Academy of Sciences, 100049, Beijing, China}




\begin{abstract}
The reverberation time is one of the most important parameters used to characterize the acoustic property of an enclosure. In real-world scenarios, it is much more convenient to estimate the reverberation time blindly from recorded speech compared to the traditional acoustic measurement techniques using professional measurement instruments. However, the recorded speech is often corrupted by noise, which has a detrimental effect on the estimation accuracy of the reverberation time.
To address this issue, this paper proposes a two-stage blind reverberation time estimation method based on noise-aware time-frequency masking.
This proposed method has a good ability to distinguish the reverberation tails from the noise, thus improving the estimation accuracy of reverberation time in noisy scenarios.   
 {The simulated and real-world acoustic experimental results show that the proposed method significantly outperforms other methods in challenging scenarios.}
\end{abstract}


\end{frontmatter}


\section{INTRODUCTION}
When sound propagates from a point to the observation point in an enclosure, it follows not only the direct path but also multiple reflections \cite{kuttruff2013room}. This physical process causes the fact that sound produced in the room will remain audible for a while after the sound source is shut off. The time taken for the sound to diminish by 60 dB at cessation is defined as the reverberation time ($T_{60}$) \cite{iso20083382}, which is mainly determined by the room geometry and the reflectivity of the surfaces \cite{iso2009acoustics}.
As one of the most important acoustical parameters in a room, $T_{60}$ can be measured using two methods in the ISO standard \cite{iso20083382}, including the interrupted noise method and the integrated impulse response method \cite{schroeder1965new}. 
However, these two methods require professional measurement instruments and expertise, taking a lot of time and efforts for large-scale measurements. Therefore, it is highly desired if $T_{60}$ can be estimated blindly, which means estimating $T_{60}$ simply from the recorded speech without any specific instruments and time-consuming acoustic measurements. 



Many blind $T_{60}$ estimation methods have been proposed for the past decades.
These approaches can be mainly classified into two categories: traditional statistical signal processing-based  methods \cite{ratnam_blind_2003,lollmann_estimation_2008,eaton_noise-robust_2013,prego_blind_2015,loellmann_single-channel_2015} and deep learning-based \cite{cox_extracting_2001,parada_evaluating_2015,xiong_joint_2015} methods.
In 2015, the ACE challenge workshop was held to evaluate the blind methods in estimating $T_{60}$ with the recorded noisy reverberant speech \cite{eaton2016estimation}.
With the rapid development of deep learning, several deep learning-based methods have been proposed after the ACE workshop and some of them have surpassed traditional methods on estimation accuracy  \cite{lee2016blind,deng2020online,srivastava2021blind}.
Most of these methods have shown satisfactory improvements in noise-free and high signal-to-noise-ratio (SNR) scenarios while the $T_{60}$ estimation in low SNR scenarios such as 0 dB are still challenging. 

However, in real-world acoustic environments, speech usually contains not only its reflections but also the uncorrelated noise. 
It has been pointed out in \cite{gaubitch2012performance} that three $T_{60}$ estimators were severely biased positively by additive white Gaussian noise. The result of ACE challenge also indicated that noise causes overestimation for most $T_{60}$ estimators. This is because the noise-like tail of reverberation, which is important for $T_{60}$ estimation, is buried in the additive noise \cite{eaton2016estimation}. 
To address this problem, some noise-robust $T_{60}$ estimation methods have been proposed \cite{lollmann_estimation_2008,eaton_noise-robust_2013,prego2015blind} to improve the estimation accuracy in low SNR scenarios.
However, most of these methods are based on traditional statistical signal processing, requiring a precise SNR estimation. The overall estimation accuracy of these methods was still not satisfactory especially in low SNR, large reverberation or non-stationary noisy scenarios, where the precise SNR estimation is difficult.
 

Recently, deep learning-based noise reduction methods have achieved great performance  \cite{reddy2021icassp,li2021icassp,liu2021know}, and they can potentially be applied to improve the robustness to noise for other tasks \cite{zhang2019robust,wang2018robust}.
To reduce the negative impact of the noise, a novel noise-aware framework based on convolutional neural networks (CNNs) is proposed in this work to improve the $T_{60}$ estimation accuracy in noisy  and reverberant environments.
The two-stage noise-aware framework consists of two sub-networks, namely the Noise Estimation Network (dubbed NE-NET) and the Reverberation time Estimation Network (dubbed RE-NET).
In the first stage, NE-NET aims to estimate the spectral magnitudes of both the noise-free reverberant speech and the noise. 
In the second stage, the estimated spectral magnitudes in the first stage together with the spectral magnitude of the recorded speech are concatenated as the input features to estimate $T_{60}$ with RE-NET.
The ideas behind the proposed method are discussed from the following two aspects. 
Firstly, the estimation of the noise-free speech can recover some reverberation details from the noisy speech, which alleviates some adverse effects of the noise.
Secondly, the SNR of each time-frequency (T-F) bin is different due to the non-stationary properties of both speech and noise and their spectral characteristics. Therefore, RE-NET can pay more attention to higher SNR T-F bins with the guidance of the estimated spectral magnitude when estimating $T_{60}$. 
To evaluate the performance of this method in various noisy scenarios, simulated and real-world experiments are carried out in this work.

{The contributions of this paper are as follows.}
(1) A novel two-stage noise-aware $T_{60}$ estimation framework is proposed to improve the robustness of estimation accuracy in noisy scenarios. 
Instead of estimating $T_{60}$ directly from noisy and reverberant speech, the proposed method uses deep neural networks to estimate and reduce the noise in the first stage. With this framework, the noise-like reverberation tails are more likely to be distinguished by deep neural networks in the second stage.
(2) Comprehensive studies considering different noise scenarios especially in low SNR  and unseen noise type conditions are investigated in this paper. 
Comparative experiments are performed to investigate the necessity of training neural networks with more complex noisy environments while most previous deep learning-based methods \cite{lee2016blind,gamper2018blind,deng2020online} only considered less than three seen noise types.
{
(3) Real-world reverberant speech samples are recorded to evaluate the performance of methods.
The reverberant speech cannot be simplified by the convolution of clean speech and RIR because the assumption of linear time invariant (LTI) system for room impulse responses could fail in real conditions \cite{kuttruff2013room}. However, most previous works evaluated their methods simply using simulated samples with the assumption of LTI system for room impulse responses, making it difficult to assess how well their methods would perform with recorded speech \cite{lee2016blind,gamper2018blind,deng2020online}.
Therefore, the proposed method are tested using recorded speech in four realistic rooms with different sizes, acoustical properties and $T_{60}$s.
}

The rest of this paper is organized as follows. In Section \ref{sec:signal model}, the problem of $T_{60}$ estimation is formulated. In Section \ref{sec:system overview}, the proposed architecture and its corresponding 
 components are discussed in detail. Section \ref{sec:experimental setup} introduces the data sets and experimental setup. Section \ref{sec:result} presents the experimental results and extensive analysis. Some conclusions are drawn in Section \ref{sec:conclu}.
 The details of the acoustic experiments are introduced in the appendix A.

\section{SIGNAL MODEL } \label{sec:signal model}
In the time domain, a noisy and reverberant speech signal is typically modeled as
\begin{equation}
\setlength{\abovedisplayskip}{2pt}
\setlength{\belowdisplayskip}{2pt}
y\left( t \right) = s\left( t \right) \ast h\left( t \right) + n\left( t \right),
\label{eq:signalmodel}
\end{equation}
where $y\left( t \right)$, $s\left( t \right)$ and $n\left( t \right)$ refer to the noisy and reverberant speech, the clean speech and the additive noise, respectively, in the time index $t$. $\ast$ denotes the convolution operation and $h\left( t \right)$ is the room impulse response from the speech source to the observation point. The noise-free reverberant speech $ x\left( t \right)=s\left( t \right) \ast h\left( t \right)$ can be viewed as the convolution of the clean speech and RIR. In realistic scenarios, $n\left( t \right)$ also contains reverberation in an enclosure but with a different RIR from $h\left( t \right)$.

With short-time Fourier transform (STFT), the formula in (\ref{eq:signalmodel}) can be transformed into
\begin{equation}
\setlength{\abovedisplayskip}{2pt}
\setlength{\belowdisplayskip}{2pt}
Y\left(k, l \right)= X\left(k, l \right) + N\left( k, l \right),
\end{equation}
where $Y\left(k, l\right)$, $X\left(k, l\right)$ and $N\left(k, l\right)$ denote the time-frequency representations of the noisy and reverberant speech, the noise-free reverberant speech and the additive noise, respectively, with the frequency bin index $k$ and the time frame index $l$.
In frequency domain,
$X\left(k, l\right)$ can be written as
\begin{equation}
	\setlength{\abovedisplayskip}{2pt}
	\setlength{\belowdisplayskip}{2pt}
	X\left(k, l \right)= S\left(k, l \right)H\left(k, l \right),
\end{equation}
where $S\left(k, l\right)$ and $H\left(k, l\right)$ denote the clean speech and the RIR in the STFT domain. 
In first stage, we aim to separate the noise-free reverberant speech and additive noise components from the single-channel recorded noisy reverberant mixture. Two Ideal Ratio Masks (IRM) \cite{liu2021know}, i.e. $\widehat{M}_{\rm speech}(k,l)$ and $\widehat{M}_{\rm noise}(k,l)$, are estimated as a speech mask and a noise mask, respectively. These two masks can be applied to estimate the spectral magnitude of reverberant speech and that of the noise by the following multiplication operation
\begin{equation}
	\setlength{\abovedisplayskip}{2pt}
	\setlength{\belowdisplayskip}{2pt}
	|\widehat{X}\left(k, l \right)|= \widehat{M}_{\rm speech}(k,l)|Y\left(k, l \right)|,
\end{equation}
and
\begin{equation}
	\setlength{\abovedisplayskip}{2pt}
	\setlength{\belowdisplayskip}{2pt}
	|\widehat{N}\left(k, l \right)|= \widehat{M}_{\rm noise}(k,l)|Y\left(k, l \right)|,
\end{equation}
where $|\widehat{X}\left(k, l \right)|$ and $|\widehat{N}\left(k, l \right)|$ denote the estimated spectral magnitude of the reverberant speech and that of the noise, respectively. For simplicity, the frequency bin index $k$ and the time frame index $l$ are omitted when no confusion arises.

\section{PROPOSED METHOD} \label{sec:system overview}


Based on the fact that the late reverberation is buried in noise especially in low SNR scenarios, the NE-Net is explicitly designed to estimate the noise-free reverberant speech and noise in this noise-aware framework.
With the noise-free reverberant speech estimation and the noise estimation in the first stage, the RE-Net can estimate the $T_{60}$ based on the prior speech and noise information in the second stage, which may alleviate some adverse effects of noise.

\subsection{Proposed Noise-aware Framework}
\begin{figure}[H]
\setlength{\abovecaptionskip}{0.235cm}
\setlength{\belowcaptionskip}{-0.1cm}
\centering
\includegraphics[width=\textwidth]{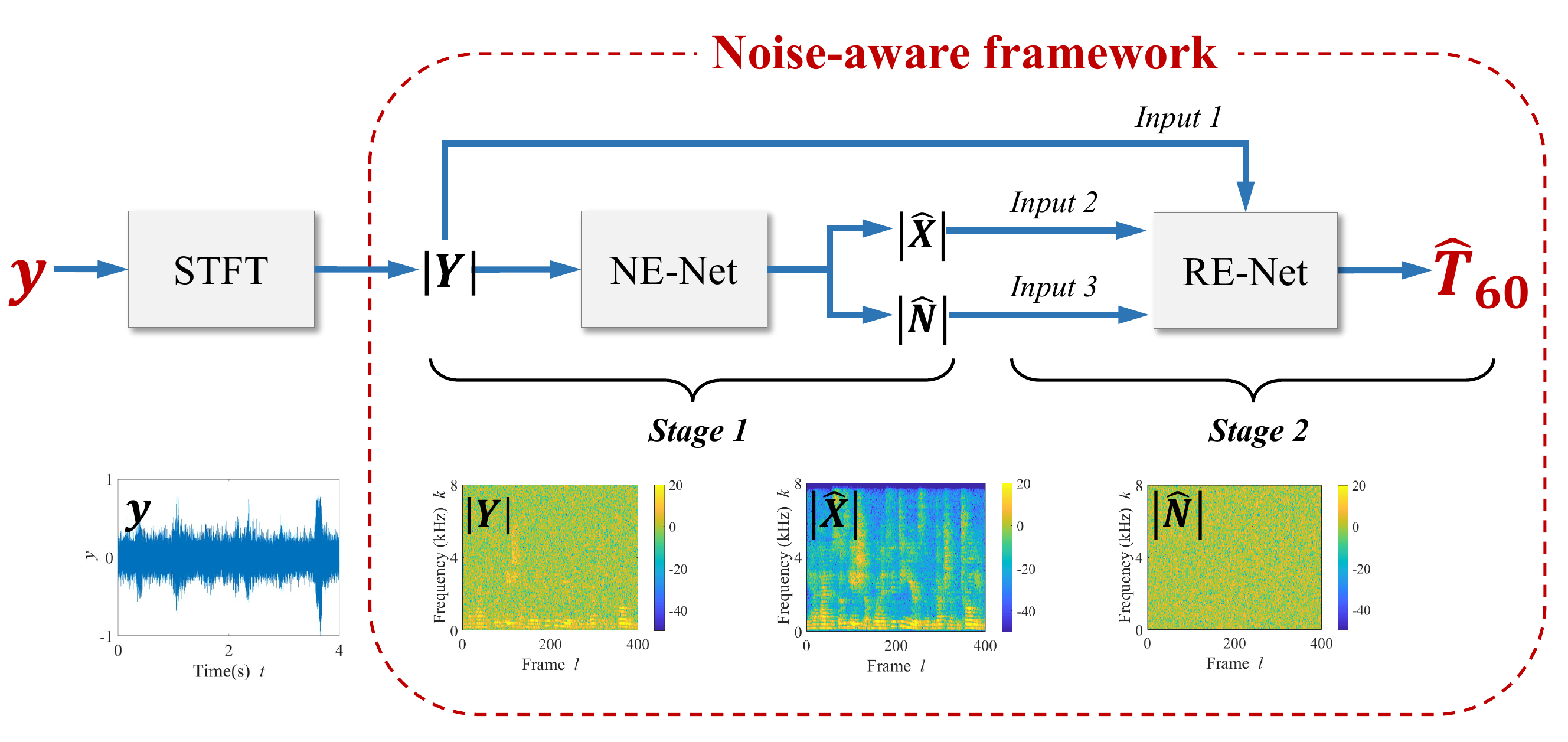}

\caption{The architecture of the proposed noise-aware framework.}
\label{fig:network}
\end{figure}

Figure \ref{fig:network} depicts the architecture of the proposed noise-aware framework. It consists of two sub-networks, namely NE-Net and RE-Net. The whole estimation process goes as follows. In the first stage, the NE-Net aims to estimate the spectral magnitude of noise-free reverberant speech and that of the noise from recorded noisy reverberant speech in STFT domain. In the second stage, the estimated spectral magnitudes together with the spectral magnitude of the original noisy reverberant speech which contains the unprocessed reverberation information are sent to RE-Net for estimating $T_{60}$.
In a nutshell, the whole forward calculation process can be formulated as:

\begin{equation}
	\setlength{\abovedisplayskip}{2pt}
	\setlength{\belowdisplayskip}{2pt}
	\left\{\widehat{M}_{\rm speech},\widehat{M}_{\rm noise} \right\}= \mathfrak{F}_1\left(|Y|;\Phi_1 \right),
\end{equation}
\begin{equation}
	\setlength{\abovedisplayskip}{2pt}
	\setlength{\belowdisplayskip}{2pt}
	|\widehat{X}|= \widehat{M}_{\rm speech}|Y|,
	|\widehat{N}|= \widehat{M}_{\rm noise}|Y|,
\end{equation}
\begin{equation}
	\setlength{\abovedisplayskip}{2pt}
	\setlength{\belowdisplayskip}{2pt}
	\widehat{T}_{60}= \mathfrak{F}_2\left(|Y|,|\widehat{X}|,|\widehat{N}|;\Phi_2 \right),
\end{equation}
where $\mathfrak{F}_1$ and $\mathfrak{F}_2$ denote mapping functions for the first stage and the second stage with parameter sets $\Phi_1$ and $\Phi_2$, respectively. 
The architecture of NE-NET resembles the U-NET proposed in \cite{ronneberger2015u} with a convolutional encoder and decoder. The RE-NET keeps the encoder part while replacing the decoder part with a fully-connected layer, which directly mapping the input features to $T_{60}$.
The modules used in the network will be introduced in the following parts.

\subsection{Detailed Sub-Networks}

\begin{figure}[H]
	\setlength{\abovecaptionskip}{0.05cm}
	\setlength{\belowcaptionskip}{-0.5cm}
	\centering
	\includegraphics[width=0.6\textwidth]{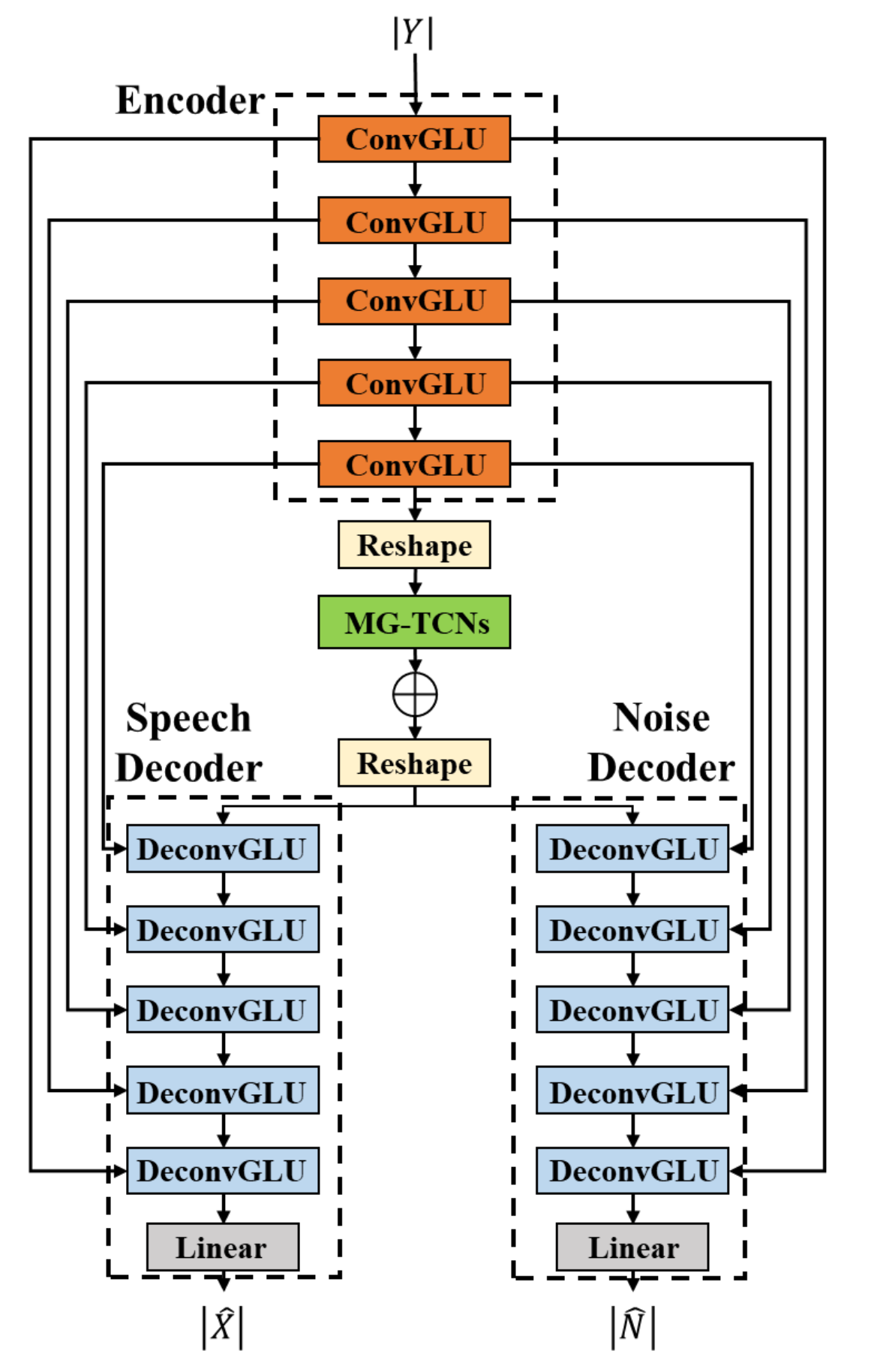}
	\caption{The detailed network architecture of NE-NET for estimating speech and noise masks. The parameter setting of each layer can be found in Table \ref{tab:detailpara}.}
	\label{fig:nedetail}
\end{figure}

In this part, the details of the two sub-networks used in the architecture will be introduced. The detailed diagram of the NE-Net is shown in Figure \ref{fig:nedetail}. 
Because the NE-NET and RE-NET have a similar network topology except that the RE-NET does not have any decoders and the number of input feature channels is three, the detailed diagram of the RE-Net is omitted for simplicity.
For NE-NET, the ReLU activation function is added after the linear layer to output a positive value for IRM. 
Each module used in the network will be discussed below.

In Figure \ref{fig:nedetail}, the encoder contains 5 ConvGLUs \cite{dauphin2017language} to compress the frequency feature dimension and the decoder contains 5 DeconvGLUs to increase the frequency feature dimension as in \cite{tan2019learning}. The encoder module and MG-TCNs \cite{li2021two} module are shared in the network while there are two different decoders for speech mask and noise mask estimation. This design is inspired by multi-task learning \cite{ruder2017overview}, which utilizes the information between different relevant tasks to achieve a better overall performance than treating tasks separately. The shared encoder module and MG-TCNs module aim to extract the features, while the two different decoders are used to map the speech mask and the noise mask, respectively.
Since speech mask and noise mask estimation are two relevant tasks \cite{wang2017joint}, the parameter sharing mechanism is expected as a regularization which may lead to a better generalization ability.

Table \ref{tab:detailpara} shows the detailed parameter setup for NE-Net.
The input size and output size of features for each layer are given in $Timesteps\times Featuresize$ format for a 2-D feature and $Channelnum\times Timesteps\times Featuresize$ format for a 3-D feature.
The layer hyper-parameters of the encoder and decoders are specified with ($Kernelsize, Strides, Outchannels$) format.
We use the kernel size of 2 $\times$ 3 ($Time \times Frequency$) as in \cite{tan2019learning} for ConvGLUs and DeconvGLUs except for 2 $\times$ 5 in the first and final layer.
Each ConvGLU and DeconvGLU block is followed by an instance normalization \cite{ulyanov2016instance} layer and a PReLU \cite{he2015delving} activation function.

As in \cite{luo2019conv}, the MG-TCNs are stacked to get a larger receptive field. In this work, 3 groups of MG-TCN each containing five MG-TCN units are used for temporal modeling. For each MG-TCN, the dilation rate of each MG-TCN unit increases exponentially from 1 to 16. The parameters are specified with ($Kernelsize, Dilationrate, Outchannels$) format in Table \ref{tab:detailpara}. We have tried several parameter settings and this configuration yields the best results. 

{
	\renewcommand\arraystretch{0.7}
	\setlength{\tabcolsep}{5pt}
	\vspace{-0.3cm}
	\begin{table}[t]
		\tiny
		\footnotesize
		\caption{Detailed parameter setup of the NE-NET. The parameter setup of RE-NET is similar to NE-NET except that the first input channel is 3 and the outchannels of the encoder and the MG-TCN unit are decreased to 10.}
		
		\begin{center}
			\begin{tabular}{c|c|c|c}
				\hline
				layer name & input size & hyperparameters & output size\\
				\hline
				conv2d\_glu\_1 & $q$ $\times$ $T$ $\times$ 161 & 2 $\times$ 5, (1, 2), 64 & 64 $\times$ $T$ $\times$ 79\\
				\hline
				conv2d\_glu\_2 & 64 $\times$ $T$ $\times$ 79 & 2 $\times$ 3, (1, 2), 64 & 64 $\times$ $T$ $\times$ 39\\
				\hline
				conv2d\_glu\_3 & 64 $\times$ $T$ $\times$ 39 & 2 $\times$ 3, (1, 2), 64 & 64 $\times$ $T$ $\times$ 19\\
				\hline
				conv2d\_glu\_4 & 64 $\times$ $T$ $\times$ 19 & 2 $\times$ 3, (1, 2), 64 & 64 $\times$ $T$ $\times$ 9\\
				\hline
				conv2d\_glu\_5 & 64 $\times$ $T$ $\times$ 9 & 2 $\times$ 3, (1, 2), 64 & 64 $\times$ $T$ $\times$ 4\\
				\hline
				reshape\_size\_2& 64 $\times$ $T$ $\times$ 4 & - & T $\times$ 256\\
				\hline
				MG\_TCNs 		&$T$ $\times$ 256 & $\begin{pmatrix}
					5 & 1 & 64\\
					5 & 2 & 64\\
					5 & 4 & 64\\
					5 & 8 & 64\\
					5 & 16 & 64\\
				\end{pmatrix}\times3$ & $T$ $\times$ 256\\
				\hline
				reshape\_size\_3 & $T$ $\times$ 256 & - & 64 $\times$ $T$ $\times$ 4\\
				\hline
				{skip\_connect\_1} &{64 $\times$ $T$ $\times$ 4} & - & {128 $\times$ $T$ $\times$ 4}\\
				\hline
				deconv2d\_glu\_1 & 128 $\times$ $T$ $\times$ 4 & 2 $\times$ 3, (1, 2), 64 & 64 $\times$ $T$ $\times$ 9\\
				\hline
				{skip\_connect\_2} &{ 64 $\times$ $T$ $\times$ 9} & - & {128 $\times$ $T$ $\times$ 9}\\
				\hline
				deconv2d\_glu\_2 & 128 $\times$ $T$ $\times$ 9 & 2 $\times$ 3, (1, 2), 64 & 64 $\times$ $T$ $\times$ 19\\
				\hline
				{skip\_connect\_3} & {64 $\times$ $T$ $\times$ 19} & - & {128 $\times$ $T$ $\times$ 19}\\
				\hline
				deconv2d\_glu\_3 & 128 $\times$ $T$ $\times$ 19 & 2 $\times$ 3, (1, 2), 64 & 64 $\times$ $T$ $\times$ 39\\
				\hline
				{skip\_connect\_4} & {64 $\times$ $T$ $\times$ 39} & - & {128 $\times$ $T$ $\times$ 39}\\
				\hline
				deconv2d\_glu\_4 & 128 $\times$ $T$ $\times$ 39 & 2 $\times$ 3,(1, 2), 64 & 64 $\times$ $T$ $\times$ 79\\
				\hline
				{skip\_connect\_5} & {64 $\times$ $T$ $\times$ 79} & - & {128 $\times$ $T$ $\times$ 79}\\
				\hline
				deconv2d\_glu\_5 & 128 $\times$ $T$ $\times$ 79 & 2 $\times$ 3, (1, 2), 1 & 1 $\times$ $T$ $\times$ 161\\
				\hline
				reshape\_size\_4 &1 $\times$ $T$ $\times$ 161 & - & $T$ $\times$ 161\\
				\hline
				linear(ReLU)	 &1 $\times$ $T$ $\times$ 161 & 161 & $T$ $\times$ 161\\
				\hline
			\end{tabular}
			\label{tab:detailpara}
		\end{center}
	\end{table}
}

\subsection{Loss Function and Optimization Strategy}

In the proposed noise-aware architecture, we train the NE-NET in the first stage and RE-NET in the second stage as mentioned previously. The trained NE-NET will be fine-turned in the second stage to get a global optimization. 
Different from most speech enhancement algorithms \cite{li2020speech,yuan2015speech}, NE-NET is trained in the first stage with a loss containing the MSE losses of both speech and noise formulated as:
\newcommand{\bignorm}[1]{\Bigl \| #1 \Bigr \|}
\begin{equation}
	\setlength{\abovedisplayskip}{2pt}
	\setlength{\belowdisplayskip}{2pt}
	L_{NE}= \alpha\Bigl \| |\widehat{X}|-|X| \Bigr \|^2_F + \left( 1-\alpha\right) \Bigl \| |\widehat{N}|-|N| \Bigr \|^2_F,
\end{equation}
where $\bignorm{\bullet}^2_F$ denotes the square of the Frobenius norm. $\alpha$ is the weighting between speech loss and noise loss.
In the second stage, we train RE-NET and refine NE-NET jointly, the total loss function can be written as:
\begin{equation}
	\setlength{\abovedisplayskip}{2pt}
	\setlength{\belowdisplayskip}{2pt}
	L= L_{RE}+\lambda L_{NE},
\end{equation}
where $\lambda $ is the weighting parameter between the two losses, and the loss function of RE-NET can be presented as:
\begin{equation}
	\setlength{\abovedisplayskip}{2pt}
	\setlength{\belowdisplayskip}{2pt}
	L_{RE}=\frac{1}{T} \sum_{l=1}^{T} \left(\widehat{T}_{60,l}-T_{60} \right)^2 ,
\end{equation}
where $l$ is the frame index. For each frame, an estimated reverberation time $\widehat{T}_{60,l}$ are outputted and all the estimated values and the ground truth value are used for calculating the MSE loss.
During the evaluation stage, only the last estimated value, i.e. the $\widehat{T}_{60}$ of the last frame will be used as the final estimated value as the estimation in the last frame can utilize the whole information of the utterance.
The values of $\alpha$ and $\lambda$ are set to 0.5 and 0.1, respectively.

\section{EXPERIMENTS} \label{sec:experimental setup}
\subsection{Datasets}
\subsubsection{Training Sets and Validation Sets}
The anechoic speech samples are taken from the WSJ0-SI84 dataset \cite{paul1992design}, which consists of 7138 utterances by 83 speakers (42 males and 41 females), to generate the dataset for training the deep neural networks. 5428 and 957 clean utterances are chosen for training and model validation, respectively.
The image-source method \cite{habets2006room} is used to generate the 2457 synthetic RIRs for cuboid environments of varying sizes and with varying absorption coefficients. We simulate 7 room sizes as specified by the ACE challenge \cite{eaton2016estimation}. The distance between the sound source and microphone is randomly chosen from the distance set $D$ $\left\lbrace 0.7m, 1m, 1.7m, 2m, 2.5m\right\rbrace$ . The angle between the sound source and microphone is randomly chosen from the angle -45\degree to 45\degree with an increment of 10\degree. The $T_{60}$ ranges from 0.2 s to 1.5 s. The ground truth of $T_{60}$ is calculated using the method in \cite{antsalo2001estimation}.
Two unseen room sizes are simulated to generate 234 synthetic RIRs for the validation set.
To alleviate the artifact of simulated RIRs, the synthetic RIRs are combined with measured RIRs from public databases, including the Openair database \cite{murphy2010openair}, the REVERB Challenge dataset \cite{kinoshita2013reverb}, and the RWCP database \cite{nakamura2000acoustical}. The multi-channel RIRs are separated into single-channel RIRs, resulting in a total of 2432 single-channel measured RIRs. 2115 and 317 measured RIRs are selected for training and validation, respectively.

To improve the robustness of $T_{60}$ estimation in noisy reverberant environments, we choose 6303 types of environmental noise from the Interspeech 2020 DNS-Challenge dataset \cite{reddy2020interspeech} as the noise set and randomly separate 5358 and 945 noise files into the training set and the validation set.
To generate a noisy reverberant speech, a randomly selected clean speech and noise are convoluted with a randomly chosen RIR, respectively. The amplitude of the noise is adjusted according to a predefined SNR randomly chosen from three different SNRs including 0 dB, 10 dB, and 20 dB. Then, the noise is added to the reverberant speech. Each utterance is truncated to four seconds. This results in a total of 63251 and 4974 noisy reverberant utterances for training and validation, respectively. The total duration of the training set is around 70 hours.

\subsubsection{Simulated Test Sets}

To evaluate the generalization ability of the proposed method in various noisy scenarios, we generate 4 different test sets including unseen speakers, seen and unseen noise types, simulated unseen acoustic environments, and realistic unseen acoustic environments.
We select 90 clean speech samples that have a duration longer than 4 s from the TIMIT dataset \cite{garofolo1993timit}. For testing the impact of speech duration on estimation accuracy in Section  \ref{sec:effect of duration}, we splice every two pieces of speech to form 45 clean speech samples with a duration longer than 8 s. The simulated RIRs are created with the same procedure as the training set except that the room sizes are different from those in the training sets. The realistic RIRs are taken from the ACE development set and evaluation set \cite{eaton2016estimation}, which contain RIRs of 7 different rooms. We separate the multichannel RIRs into single-channel RIRs for data augmentation. The 200 seen noise processes are randomly chosen from the training set while the 14 noise processes taken from NOISEX92 \cite{varga1993assessment} are used as unseen noise. Five SNRs are set, namely 0 dB, 5 dB, 10 dB, 15 dB and 20 dB.

For the test set 1, simulated RIRs and seen noise processes are used. For the test set 2, realistic RIRs and seen noise processes are used. For the test set 3, simulated RIRs and unseen noise processes are used. For the test set 4, realistic RIRs and unseen noise processes are used.
Table \ref{table:datasets} summarises the data sets used in this work.

\begin{table}[htbp]
	\caption{An overview of training, validation and test sets.}
	\centering
	\footnotesize
	\begin{tabular}{c c c c}
		\hline
		Type &RIR data &noise &samples \\
		\hline
		training &2457 synthetic and 2432 measured RIRs &5358 DNS noise &63251\\
		validation &234 synthetic and 317 measured RIRs &945 DNS noise &4974\\
		testing1 &2437 synthetic RIRs &200 seen noise &1000\\
		testing2 &700 real RIRs &200 seen noise &1000\\
		testing3 &2437 synthetic RIRs &14 unseen noise &1000\\
		testing4 &700 real RIRs &14 unseen noise &1000\\
		\hline
	\end{tabular}
	\label{table:datasets}
\end{table}

\subsection{Parameter Configurations}

All the utterances are sampled at 16 kHz and split into chunks of 4 seconds for training stability as in \cite{gamper2018blind}. For STFT configuration, the 20 ms Hanning window is used with $50\%$, i.e. 10ms, overlap. 320-point FFT is used, resulting in 161-D spectral features.
Adam \cite{kingma2014adam} with a 0.001 learning rate is used in the first stage. In the second stage, NE-NET is fine-tuned with a learning rate of 0.0001, and the learning rate for RE-NET is set to 0.001. For better convergence, the learning rate will be halved if the validation loss does not decrease during 3 consecutive epochs. The total epochs for the network training are 120, 60 for the first stage and 60 for the second stage. The batch size is set to 8 at the utterance level.

\subsection{Baselines}

To evaluate the performance of the proposed method, three baselines, namely the CNN proposed in \cite{gamper2018blind}, RE-NET, and masking method, are implemented for comparison. 
RE-NET means only the RE-NET in the noise-aware architecture is used to receive the noisy reverberant speech and estimate $T_{60}$ directly. RE-NET is implemented to test the effectiveness of the NE-NET.
The masking method is also a two-stage method as the noise-aware method. However, the masking method only outputs a speech mask in the first stage and  estimates $T_{60}$ in the second stage. The masking method is proposed to show whether using estimated noise as the additional input in the second stage can improve the performance of the network.

The four methods can be classified into two categories, namely, direct method and denoise-based indirect method. The CNN and RE-NET belong to the direct method, which directly estimates $T_{60}$ using the feature extracted from the recorded noisy reverberant speech. The masking method and noise-aware method belong to the denoise-based indirect method, which contains two stages, i.e. the denoise stage and the estimation stage. Note that the noise-aware method not only extracts the reverberant speech but also estimates the noise as inputs.

\subsection{Evaluation Metrics}
\subsubsection{Estimation Errors}
The estimation error is defined as the difference between the estimated value and the ground truth, which can be formulated as
\begin{equation}
	\setlength{\abovedisplayskip}{2pt}
	\setlength{\belowdisplayskip}{2pt}
	e_{T_{60}} =T_{60}-\widehat T_{60},
\end{equation}
with $\widehat T_{60}$ and $T_{60}$ denoting the estimated $T_{60}$ and the ground truth, respectively. For a batch of $n$ samples, we define the root mean squared error (RMSE) as
\begin{equation}
	\setlength{\abovedisplayskip}{2pt}
	\setlength{\belowdisplayskip}{2pt}
	RMSE_{T_{60}} =\sqrt{{\frac {1}{n}}\sum\limits_{i=1}^{n}(T_{60,i}-{\widehat T_{60,i} })^{2}},
\end{equation}
Where $i$ is the index of each sample.

\subsubsection{Pearson Correlation Coefficient}
Estimation error and RMSE cannot fully determine how well the estimator performs. 
This is because they are not normalized by the ground truth of the reverberation time $T_{60}$.
Therefore, Pearson correlation coefficient denoted as $\rho$ is used as another evaluation metric.
The better the algorithm is, the closer $\rho$ to 1 is.
The formulation of Pearson correlation coefficient $\rho$ for $n$ measurement samples can be written as

\begin{equation}
	\setlength{\abovedisplayskip}{2pt}
	\setlength{\belowdisplayskip}{2pt}
	\rho_{T_{60},\widehat T_{60}}={\frac {\sum\limits_{i=1}^{n}(T_{60,i}-{\bar T_{60}})(\widehat T_{60,i}-  \bar{\widehat{T}}_{60} )} {{\sqrt{\sum\limits_{i=1}^{n}(T_{60,i}-\bar{T}_{60})^{2}}}{\sqrt {\sum\limits_{i=1}^{n}(\widehat T_{60,i}-\bar{\widehat{T}}_{60})^{2}}}}},
\end{equation}
where $\bar{\widehat{T}}_{60}$ and $\bar T_{60}$ denote the mean values of estimated $T_{60}$ and its ground truth, respectively.

\section{RESULTS AND DISCUSSION} \label{sec:result}

\subsection{ {Effect Of SNR on Estimation Accuracy}} 

{We first evaluate the performance of all methods on the four test sets to investigate the estimation accuracy in different SNR scenarios.}
Table \ref{table:all-results} shows a comprehensive comparison of the proposed methods and the CNN method proposed in \cite{gamper2018blind}. 
From this table, the RMSE and pearson correlation coefficient are highly correlated in most cases.
In general, the estimation accuracy of each method increases with SNR, which implies that noise has detrimental effects on the estimation results. This result agrees with the previous observation of other researchers \cite{eaton2016estimation,gaubitch2012performance}.

From the results, the proposed noise-aware method and masking method outperforms CNN and RE-NET estimation methods in most cases, especially in low SNR situations. This observation indicates that the first stage, i.e. the denoising stage, can alleviate some negative effects of the noise.
However, for some high SNR, i.e. 15 dB and 20 dB, and simulated RIR situations where the estimation environment is relatively simple, the denoising stage has little improvement on estimation accuracy.
Then we focus on low SNR such as 0 dB and 5 dB situations, the noise-aware method and the masking method outperform the RE-NET estimation method consistently. The relative improvement decreases with SNR, which illustrates the superiority of the denoise-based methods on low SNR conditions. Between these two denoise-based methods, the noise-aware method outperforms the masking method in most cases, which means that the estimated noise information is a useful feature in the reverberation time estimation stage.

\renewcommand\arraystretch{0.8}
\newcolumntype{V}{!{\vrule width 2pt}}
\begin{table*}[t]
	\caption{Experimental results of four methods in different noisy reverberant scenarios. Bold type indicates the best result for each case. }
	\vspace{0.1cm}
	\centering
	\LARGE
	\resizebox{\textwidth}{!}{
		
		\begin{tabular}{cVcVccccccVcccccc}
			\Xhline{1pt}
			\Xhline{2pt}
			\multicolumn{2}{cV}{{Metrics}}  & \multicolumn{6}{cV}{{RMSE (ms)}}	& \multicolumn{6}{c}{{$\rho$}}  \\
			\Xhline{2pt}
			\multicolumn{2}{cV}{{SNR (dB)}} & {0}  &{5} &{10} &{15} &{20} &{Avg.}
			& {0}  &{5} &{10} &{15} &{20} &{Avg.}
			\\
			\Xhline{2pt}
			\multirow{6}*{seen+simu}
			&  CNN     & 298& 253& 234& 211& 219& 245& 0.763& 0.826& 0.849& 0.886& 0.857& 0.835\\
			&  RE-NET  & 272& 205& 202& 146& \textbf{173}& 204& 0.799& 0.884& 0.881& 0.944& \textbf{0.908}& 0.88\\
			&  MASKING & 213& 179& 188& 152& 183& 184& 0.892& 0.926& \textbf{0.904}& 0.944& 0.897& 0.911\\
			&  NOISE-AWARE & \textbf{190}& \textbf{168}& \textbf{186}& \textbf{142}& 183& \textbf{175}& \textbf{0.913}&\textbf{ 0.927}& 0.903& \textbf{0.951}& 0.899& \textbf{0.918}\\
			\Xhline{2pt}
			\multirow{6}*{seen+real}
			&  CNN     & 249& 219& 203& 205& 211& 218& 0.616& 0.762& 0.825& 0.875& 0.854& 0.794\\
			&  RE-NET  & 260& 202& 195& 177& 166& 202& 0.582& 0.825& 0.832& 0.904& 0.905& 0.822\\
			&  MASKING & 211& 186& 161& 168& 158& 178& 0.773& 0.872& 0.906& 0.917& 0.919& 0.882\\
			&  NOISE-AWARE & \textbf{197}& \textbf{157}& \textbf{151}& \textbf{141}& \textbf{149}& \textbf{160}& \textbf{0.792}& \textbf{0.916}& \textbf{0.918}& \textbf{0.945}& \textbf{0.94}& \textbf{0.907}\\
			\Xhline{2pt}
			\multirow{6}*{unseen+simu}
			&  CNN     & 274& 239& 203& 214& 240& 236& 0.791& 0.832& 0.901& 0.874& 0.871& 0.853\\
			&  RE-NET  & 213& 172& 172& \textbf{143}& \textbf{162}& 174& 0.872& 0.909& 0.916& 0.94& \textbf{0.939}& 0.915\\
			&  MASKING & 192& 157& 162& 147& 168& 166& 0.903& 0.927& 0.934& 0.938& 0.935& 0.927\\
			&  NOISE-AWARE& \textbf{180}& \textbf{148}& \textbf{152}& 147& 177& \textbf{162}&\textbf{ 0.913}& \textbf{0.933}& \textbf{0.939}& \textbf{0.941}& 0.929& \textbf{0.93}\\
			\Xhline{1pt}
			\multirow{6}*{unseen+real}
			&  CNN     & 243& 200& 193& 183& 188& 203& 0.698& 0.832& 0.876& 0.887& 0.895& 0.833\\
			&  RE-NET  & 257& 216& 214& 181& 166& 209& 0.683& 0.814& 0.843& 0.883& 0.915& 0.831\\
			&  MASKING & 215& 192& 159& 153& 148& 176& 0.801& 0.875& 0.929& 0.923& 0.939& 0.894\\
			&  NOISE-AWARE & \textbf{203}& \textbf{175}& \textbf{145}& \textbf{142}& \textbf{129}& \textbf{161}& \textbf{0.812}& \textbf{0.883}& \textbf{0.93}& \textbf{0.93}& \textbf{0.948}& \textbf{0.901}\\
			\Xhline{2pt}
	\end{tabular}}
	\label{table:all-results}
\end{table*}

Next, we compare the differences of all methods among four test sets. By adding measured RIRs in the training stage, the discrepancy of estimation results between simulated and realistic RIR is decreased. 
For some situations, the models have better performances on unseen noise than seen noise. This can be attributed to the great generalization ability on different noise. A large number of noise processes with different SNRs are used and the noise types vary greatly in the training set. Therefore, the models become robust to unseen noise scenarios.



\begin{figure}[t]
	\setlength{\abovecaptionskip}{0.235cm}
	\setlength{\belowcaptionskip}{-0.1cm}
	\centering
	\centerline{\includegraphics[width= 0.7\columnwidth]{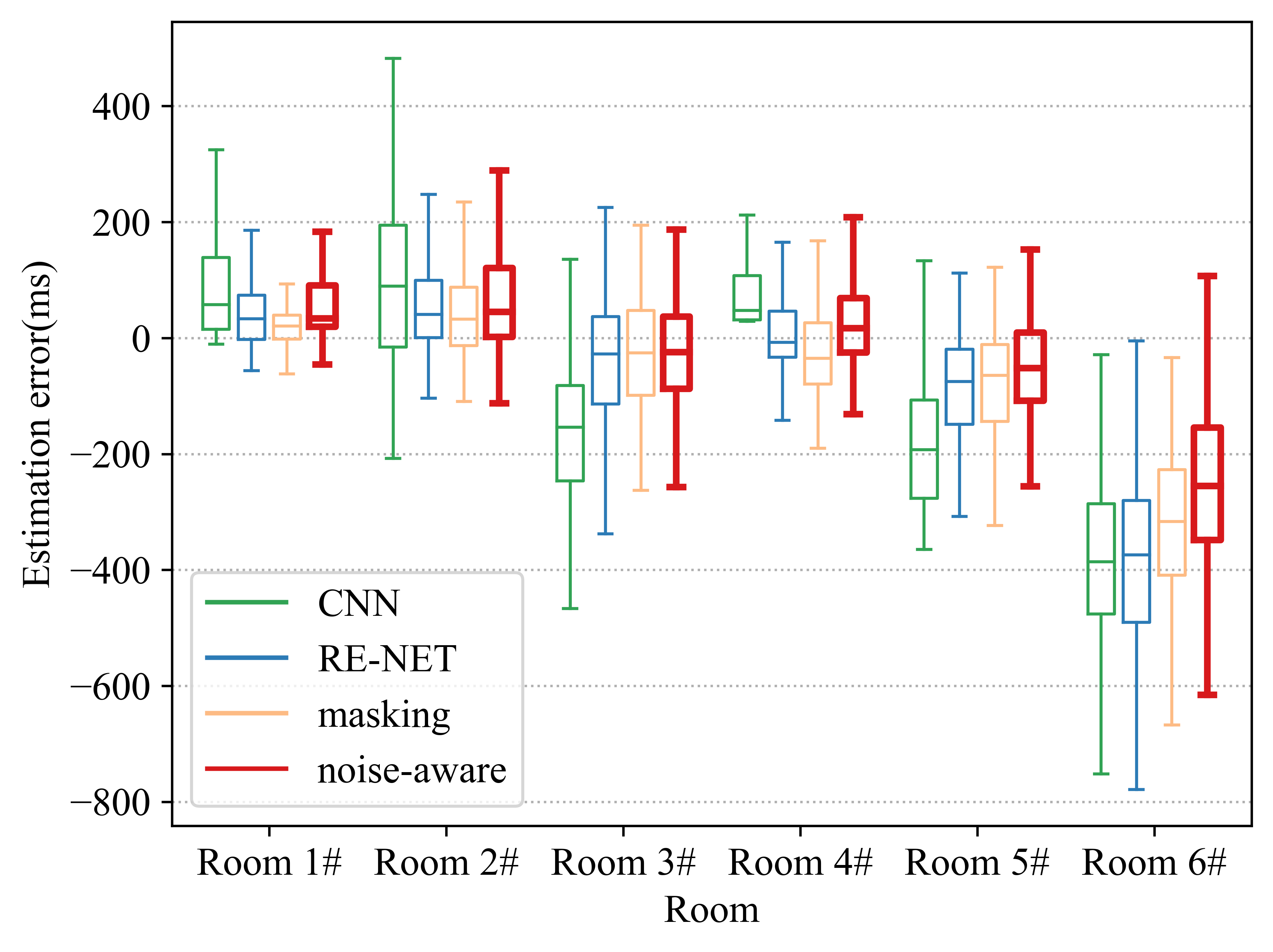}}
	\caption{ Estimation errors of proposed methods and baselines for each room. Each box displays the data based on a five-number summary: the minimum, the maximum, the median, the first and third quartiles. The notch inside the box is the median, the edges of the box are the first and third quartiles, respectively. The ends of the whiskers show the minimum and the maximum, respectively. The details of the room configuration are shown in Table \ref{table:room config}. }
	\label{figure:boxplot_eachroom}
	\vspace{0cm}
\end{figure}

\subsection{ { Effect of Reverberation Time on Estimation Accuracy}}

To investigate the estimation performance given specific $T_{60}$s, we select three simulated RIRs and three realistic RIRs with the small, medium and large $T_{60}$, respectively. The details of the room configuration are shown in Table \ref{table:room config}. 
The estimation results of the four algorithms can be seen in Figure \ref{figure:boxplot_eachroom}.
In general, the noise-aware method has a better performance than baselines. 
For the simulated rooms, i.e. room1 to room3, most models overestimate $T_{60}$ in small and medium $T_{60}$ situations while underestimating $T_{60}$ in large $T_{60}$ situations. The denoise-based methods have a similar performance while the direct method and CNN have relatively larger variances. The CNN method greatly underestimates $T_{60}$ in large $T_{60}$ environments.
In the realistic RIRs scenarios, i.e. room4 to room6, models have worse performances than in the simulated RIRs situations, especially in median and large $T_{60}$ situations. We infer that this observation results mainly from the following reasons. 
\begin{table}[t]
	\caption{An overview of room configuration given specific rooms.}
	\centering
	\footnotesize
	\begin{tabular}{c c c c c c c c}
		\hline
		Name &L(m) &W(m) &H(m) &Vol.(m$^{3}$) &RIR type &$T_{60}$(s) \\
		\hline
		Room 1\# &3.85 &5.33 &3.86 &79.6 &simulated &0.416\\
		Room 2\# &3.81 &4.65 &2.62 &46.5 &simulated &0.751\\
		Room 3\# &4.48 &6.96 &3.12 &97.5 &simulated &1.301\\
		Room 4\# &3.32 &4.83 &2.95 &47.3 &realistic &0.377\\
		Room 5\# &4.47 &5.13 &3.18 &72.9 &realistic &0.782\\
		Room 6\# &13.6 &9.29 &2.94 &370 &realistic &1.316\\
		\hline
	\end{tabular}
	\label{table:room config}
\end{table}
(1) Imbalanced realistic RIR data. As mentioned in the datasets preparation, we use the noisy reverberant speech produced by convoluting clean speech utterances with RIRs from online realistic RIRs datasets to alleviate the artifact of the simulated RIRs generated by the image-source method. However, $T_{60}$ of the realistic RIRs mainly ranges from 0.3 s to 0.8 s, which results in unbalanced $T_{60}$ labels. Using these RIRs may affect the generalization ability of the model in large $T_{60}$ situations.
(2) Noise effect. The noise has a similar acoustic characteristic with late reverberation. Late reverberation is essential for estimating $T_{60}$, especially in large $T_{60}$ scenarios. However, late reverberation is often buried in the noise in low SNR situations so that the neural network has difficulty in distinguishing late reverberation from noise.
(3) Effect of sequence modeling method. 
In a larger $T_{60}$ scenario, more context information should be utilized to estimate $T_{60}$, which increases the difficulty in sequence modeling. Therefore, the limitation of long context modeling may also affect the accuracy of estimation.

\begin{figure}[t]
	\setlength{\abovecaptionskip}{0.235cm}
	\setlength{\belowcaptionskip}{-0.1cm}
	\centering
	\centerline{\includegraphics[width= 0.7\columnwidth]{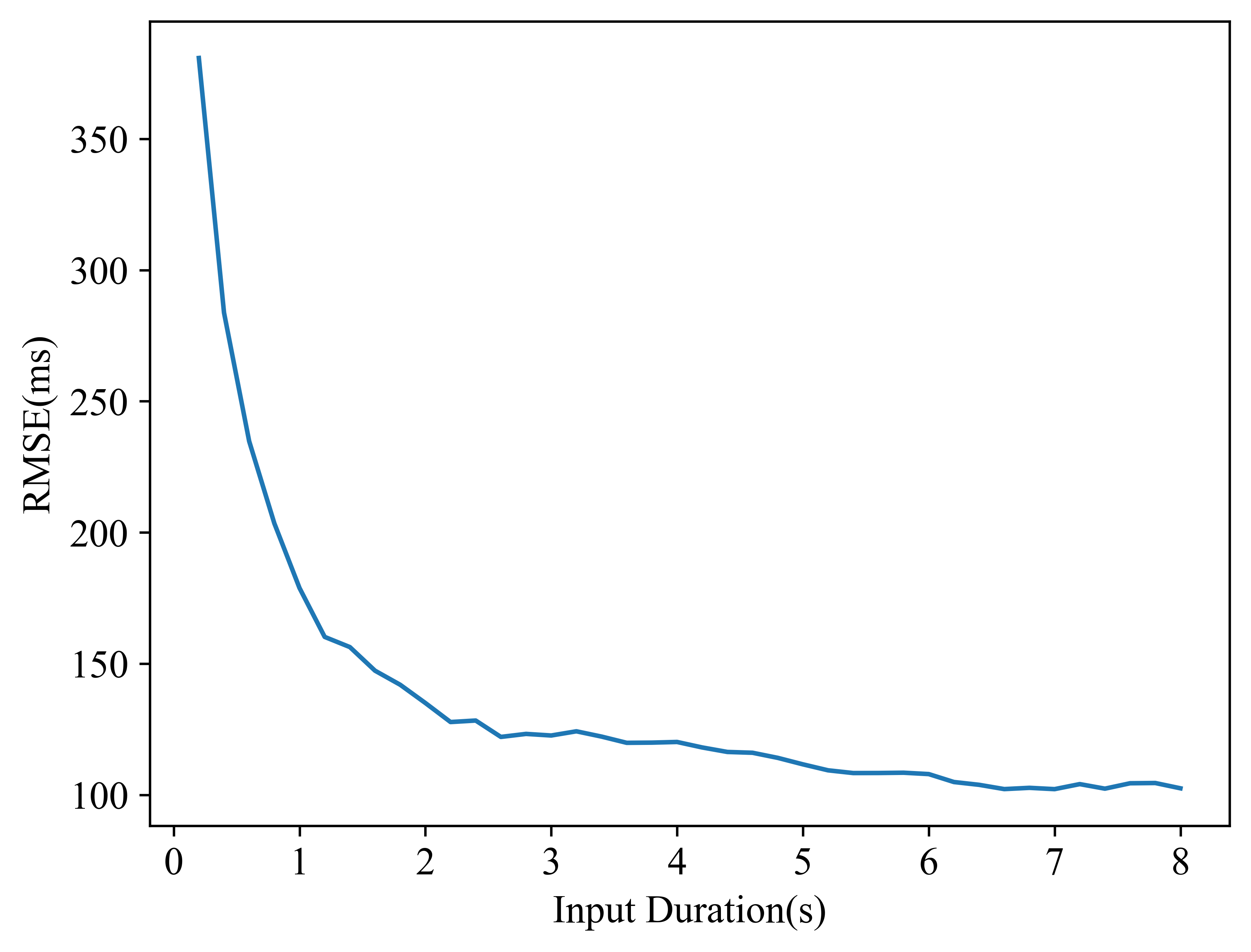}}
	\caption{ RMSE of the noise-aware algorithm on four test sets over different input duration lengths. }
	\label{figure:boxplot_errorwithtime}
	\vspace{0cm}
\end{figure}

\subsection{Effect of Input Speech Duration on Estimation Accuracy}\label{sec:effect of duration}

Since the proposed algorithm outputs one estimation for each frame, the length of the input speech is variable. 
This part studies the impact of speech duration on estimating the reverberation time.
The proposed model estimates $T_{60}$ using noisy reverberant speech samples with different durations. The length of input samples ranges from 0.2 s to 8 s with an increment of 0.2 s, resulting in 40 estimations for a 8 s long speech. Figure \ref{figure:boxplot_errorwithtime} shows the averaged RMSE of the noise-aware algorithm over the input sample length. It shows that the RMSE decays exponentially with the increase of input duration, which indicates that enough context information (about 3 s) is essential for $T_{60}$ estimation and longer input duration increases the robustness of estimation.

\subsection{Effect of Training Set on Estimation Accuracy}\label{sec:effect of noise types}

In previous sections, the training set contains various complicated noise types to ensure that the methods perform well in unseen noisy environments. However, in most previous studies \cite{gamper2018blind,deng2020online}, only a limited number of noise types were considered. To investigate the importance of increased noise types, we also train the proposed model and all baselines using a training set only containing three noise types used in the ACE challenge \cite{eaton2016estimation}, i.e. babble noise, ambient noise and fan noise, as \cite{gamper2018blind,deng2020online}.
The construction of training set is the same as the previous one except that only three noise types are used.
Then, the performance of all models is evaluated on a test set containing seen noise types (single-channel ACE Eval  \cite{eaton2016estimation}) and another one containing 14 unseen noise types (test set 4), respectively.
The experimental results are shown in Table \ref{table:3noiseresults}.
\renewcommand\arraystretch{0.8}
\newcolumntype{V}{!{\vrule width 2pt}}
\begin{table*}[t]
	\caption{Experimental results of four models trained with 3 noises (babble, ambient, fan) as \cite{gamper2018blind,deng2020online} and 5358 noises from DNS-Challenge noise corpus \cite{reddy2020interspeech} shown in Table \ref{table:all-results}. Bold type indicates the best result for each case. }
	\vspace{0.1cm}
	\centering
	\LARGE
	\resizebox{0.95\textwidth}{!}{
		
		\begin{tabular}{cVcVcVcVc}
			
			\Xhline{2pt}
			\multicolumn{2}{cV}{{Metrics}} & \multicolumn{1}{cV}{{Training noise}} & \multicolumn{1}{cV}{{RMSE (ms)}}	& \multicolumn{1}{c}{{$\qquad\rho\qquad$}}  \\
			
			\Xhline{2pt}
			\multirow{6}{3in}[10pt]{{ACE Eval \cite{eaton2016estimation}
					\qquad\qquad\qquad\qquad (3 seen noise types)}}
			&  CNN     & 3 types & 173 & 0.879\\
			&  RE-NET  & 3 types & 192 & 0.845\\
			&  MASKING & 3 types & \textbf{168} & 0.903\\
			&  NOISE-AWARE & 3 types & 175 & \textbf{0.904}\\
			\Xhline{2pt}
			\multirow{6}{3in}[10pt]{{Test set 4 
					\qquad\qquad\qquad\qquad (14 unseen noise types)}}
			&  CNN     & 3 types & 215 & 0.753\\
			&  RE-NET  & 3 types & 221 & 0.764\\
			&  MASKING & 3 types & \textbf{192} & \textbf{0.819}\\
			&  NOISE-AWARE & 3 types & 207 & 0.788\\
			\Xhline{2pt}
			\multirow{6}{3in}[10pt]{{Test set 2 
					\qquad\qquad\qquad\qquad (200 seen noise types)}}
			&  CNN     & 5358 types & 218 & 0.794\\
			&  RE-NET  & 5358 types & 202 & 0.822\\
			&  MASKING & 5358 types & 178 & 0.882\\
			&  NOISE-AWARE & 5358 types & \textbf{160} & \textbf{0.907}\\
			\Xhline{2pt}
			\multirow{6}{3in}[10pt]{{Test set 4 
					\qquad\qquad\qquad\qquad (14 unseen noise types)}}
			&  CNN     & 5358 types & 203 & 0.833\\
			&  RE-NET  & 5358 types & 209 & 0.831\\
			&  MASKING & 5358 types & 176 & 0.894\\
			&  NOISE-AWARE & 5358 types & \textbf{161} & \textbf{0.901}\\
			\Xhline{2pt}
		\end{tabular}
	}
	\label{table:3noiseresults}
\end{table*}

From the results, it shows that the models trained with 3 noises perform well on seen noise while their performances degrade greatly on unseen noise, which indicates the training the deep learning models with only 3 noise types as previous works \cite{gamper2018blind,deng2020online} results in poor robustness to unseen noise types.
As a contrast, the models trained with 5358 noises from DNS-Challenge noise corpus perform well on both seen noise and unseen noise.
Thus, it is necessary to train the models with various noise types.
For unseen noise types, the noise-aware method has worse performance than the masking method when trained with 3 noises. This means that the ability of noise estimator is limited with only 3 noise types used in the training stage. 

\subsection{{ Real-world Testing}\label{sec:Real_test}}

The reverberant speech was generated by the convolution of clean speech and RIR in simulated experiments with the assumption that room is a LTI system, and most previous works \cite{lee2016blind,gamper2018blind,deng2020online} evaluated their methods only using reverberant speech generated by this procedure. 
However, in real-world scenarios, this assumptions often falls as acoustic transmission in a room is a time-varying and non-linear process \cite{kuttruff2013room}. 
To further verify the proposed method, acoustic experiments considering real-world scenarios are investigated.
Speech and noise were recorded in four rooms with different $T_{60}$s
, including one office, two meeting rooms and one sound insulation room. 
Sizes and mean $T_{60}$s of the rooms are shown in Table \ref{table:real_room_config}. Details of the experimental procedure are introduced in Appendix A. 
The experimental results of four methods in real-world scenarios are shown in Table \ref{table:realdata-results}. 
The real-world experiment results are similar to the simulated results, which indicate that the performances of methods decrease with the appearance of additive noise. The masking and noise-aware methods have better performances especially in low SNR scenarios, showing that the denoising stage have improved the performance of methods in noisy conditions. 
A two-way analysis of variance (ANOVA) is conducted on the RMSE among different methods and SNRs. 
The results show that method [$F(3, 11996) = 262.210, p < 0.01$] and SNR [$F(4, 11995) = 89.285, p < 0.01$] have a significant effect on the RMSEs.
The interaction of method and SNR has no significant effect on the RMSEs [$F(12, 11980) = 1.032, p = 0.415$].
{The results of Fisher LSD post hoc test for the RMSEs between five methods under all SNRs are shown in Table \ref{table:ANOVA-method}.
The results show that all methods except masking and noise-aware have a significant difference in between, which indicates that the denoising stage has significantly improved the RMSE in noisy scenarios ($p < 0.01$) while the noise estimation has no significant effect on the RMSE ($p = 0.051$). 

\begin{table}[t]
	\caption{An overview of measured room dimensions and mean full-band $T_{60}$.}
	\centering
	\footnotesize
	\begin{tabular}{c c c c c c c c}
		\hline
		Name &L(m) &W(m) &H(m) &Vol.(m$^{3}$) & mean $T_{60}$(s) \\
		\hline
		Room 1\# &6.16 &4.72 &2.80 &81.41  &0.324\\      
		Room 2\# &12.42 &6.93 &2.67 &229.81  &0.822\\
		Room 3\# &6.20 &4.66 &2.79 &80.60  &0.838\\
		Room 4\# &5.20 &4.26 &3.65 &80.85  &1.512\\
		\hline
	\end{tabular}
	\label{table:real_room_config}
\end{table}

\begin{table*}[t]
	\caption{Experimental results of four methods in real-world noisy reverberant scenarios. Bold type indicates the best result for each case. }
	\vspace{0.1cm}
	\centering
	\LARGE
	\resizebox{0.95\textwidth}{!}{
		
		\begin{tabular}{cVcVccccccVcccccc}
			\Xhline{1pt}
			\Xhline{2pt}
			\multicolumn{2}{cV}{{Metrics}}  & \multicolumn{6}{cV}{{RMSE (ms)}}	& \multicolumn{6}{c}{{$\rho$}}  \\
			\Xhline{2pt}
			\multicolumn{2}{cV}{{SNR (dB)}} & {0}  &{5} &{10} &{15} &{20} &{Avg.}
			& {0}  &{5} &{10} &{15} &{20} &{Avg.}
			\\
			\Xhline{2pt}			
			\multicolumn{2}{cV}{{CNN \cite{gamper2018blind}} }     & 332& 283& 253& 236& 233& 267& 0.661& 0.771& 0.836& 0.853& 0.864& 0.797\\
			\multicolumn{2}{cV}{RE-NET}  &272& 248& 211& 181& 169& 216& 0.821& 0.868& 0.927& 0.952& 0.965& 0.907\\
			\multicolumn{2}{cV}{MASKING} &212& 176& 159& 148& 148& 168& 0.894& 0.931& 0.946& 0.960& 0.964& 0.939\\
			\multicolumn{2}{cV}{NOISE-AWARE} &\textbf{193}& \textbf{159}& \textbf{146}& \textbf{137}& \textbf{135}& \textbf{154}& \textbf{0.904}& \textbf{0.939}& \textbf{0.951}& \textbf{0.961}&\textbf{ 0.966}& \textbf{0.944}\\			
			\Xhline{2pt}
	\end{tabular}}
	\label{table:realdata-results}
\end{table*}

\begin{table}[t]
	\caption{Results of Fisher LSD post hoc test for the RMSE between four methods. A confidence interval of $95\%$ is adopted and significant effects ($p < 0.05$) are given in boldface. The Roman number indicates different methods. M = mean; SD = standard deviation; RMSE = root mean squared error. }
	\centering
	\footnotesize
	\begin{tabular}{c c c c c }
		\hline
		\multirow{2}{*}{Method} & {RMSE (ms)} & \multicolumn{3}{c}{$p$ value} \\
		\cline{3-5}
			&M $\pm$ SD	& \uppercase\expandafter{\romannumeral1} & \uppercase\expandafter{\romannumeral2} & \uppercase\expandafter{\romannumeral3} \\
		\hline
		CNN \cite{gamper2018blind} (\uppercase\expandafter{\romannumeral1}) & 206 $\pm$ 174  &   &   &   \\      
		RE-NET (\uppercase\expandafter{\romannumeral2}) 	 &   149 $\pm$ 160 & $\mathbf{<0.01}$ &   &   \\
		MASKING (\uppercase\expandafter{\romannumeral3}) 	 &   122 $\pm$ 118 & $\mathbf{<0.01}$ & $\mathbf{<0.01}$ &   \\
		NOISE-AWARE (\uppercase\expandafter{\romannumeral4})  &  115 $\pm$ 104 & $\mathbf{<0.01}$ & $\mathbf{<0.01}$ & 0.051 \\
		\hline
	\end{tabular}
	\label{table:ANOVA-method}
\end{table}

\begin{figure}[t]
	\setlength{\abovecaptionskip}{0.235cm}
	\setlength{\belowcaptionskip}{-0.1cm}
	\centering
	\centerline{\includegraphics[width= 0.7\columnwidth]{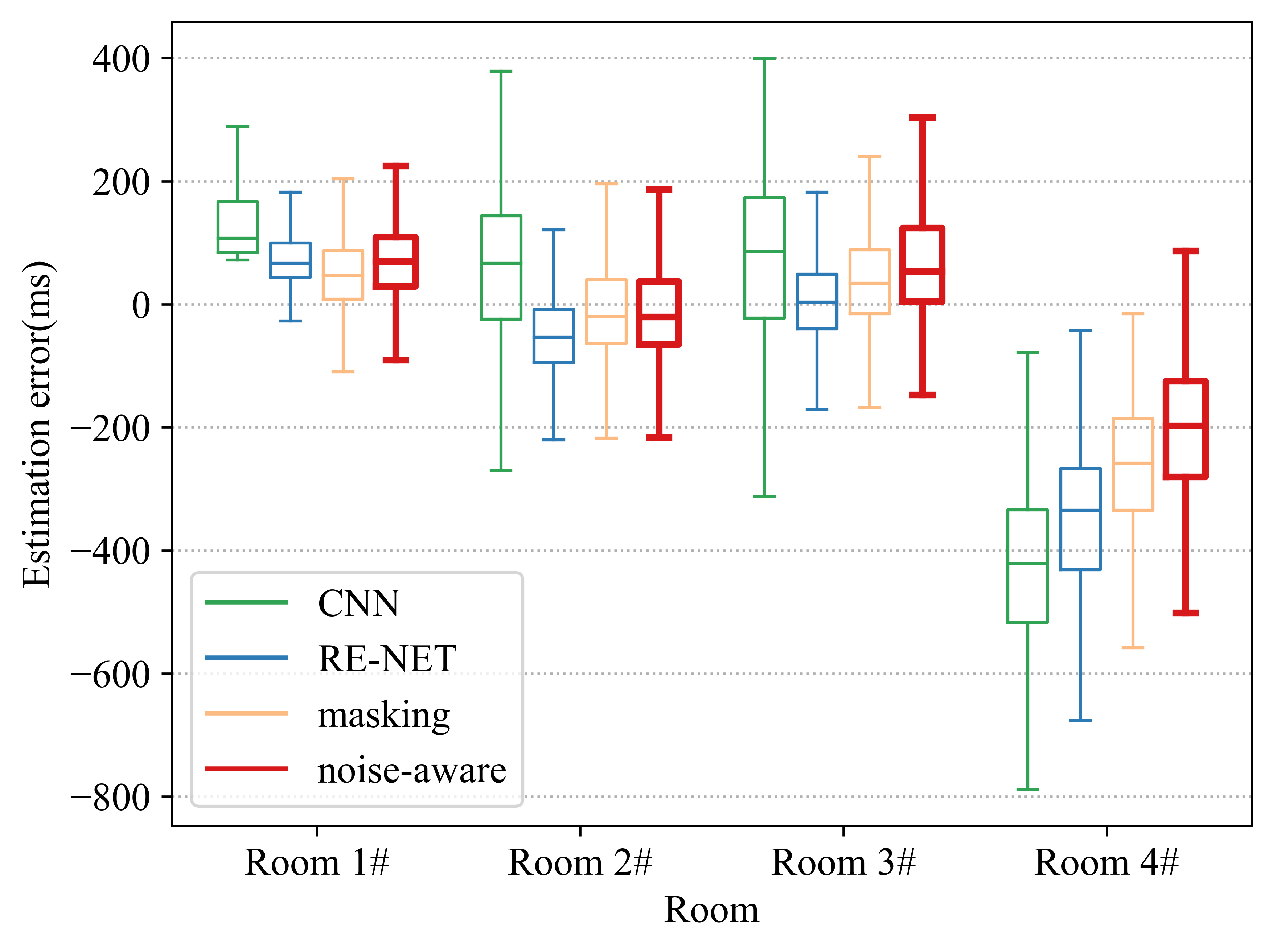}}
	\caption{ Estimation errors of proposed methods and baselines for each room in real-world scenarios. Each box displays the data based on a five-number summary: the minimum, the maximum, the median, the first and third quartiles. The notch inside the box is the median, the edges of the box are the first and third quartiles, respectively. The ends of the whiskers show the minimum and the maximum, respectively. The details of the room configuration are shown in Table \ref{table:real_room_config}. }
	\label{figure:boxplot_eachroom_real}
	\vspace{0cm}
\end{figure}

\begin{table}[t]
	\caption{Results of Fisher LSD post hoc test for the RMSE between four methods under different $T_{60}$s. A confidence interval of $95\%$ is adopted and significant effects ($p < 0.05$) are given in boldface. The Roman number indicates different methods. M = mean; SD = standard deviation; RMSE = root mean squared error. }
	\centering
	\footnotesize
	\begin{tabular}{c| c c c c c }
		\hline
		\multirow{2}{*}{$T_{60}$ (s)} & \multirow{2}{*}{Method} & {RMSE (ms)} & \multicolumn{3}{c}{$p$ value} \\
		\cline{4-6}
		&	&M $\pm$ SD	& \uppercase\expandafter{\romannumeral1} & \uppercase\expandafter{\romannumeral2} & \uppercase\expandafter{\romannumeral3} \\
		\hline
		\multirow{4}{*}{0.32} &CNN \cite{gamper2018blind} (\uppercase\expandafter{\romannumeral1}) & 137 $\pm$ 71  &   &   &   \\      
		&RE-NET (\uppercase\expandafter{\romannumeral2}) 	 &   89 $\pm$ 80 & $\mathbf{<0.01}$ &   &   \\
		&MASKING (\uppercase\expandafter{\romannumeral3}) 	 &   79 $\pm$ 90 & $\mathbf{<0.01}$ & 0.052 &   \\
		&NOISE-AWARE (\uppercase\expandafter{\romannumeral4})  &  88 $\pm$ 83 & $\mathbf{<0.01}$ & 0.967 & 0.057 \\
		\hline
		\multirow{4}{*}{0.82-0.83} &CNN \cite{gamper2018blind} (\uppercase\expandafter{\romannumeral1}) & 123 $\pm$ 85  &   &   &   \\      
		&RE-NET (\uppercase\expandafter{\romannumeral2}) 	 &   71 $\pm$ 69 & $\mathbf{<0.01}$ &   &   \\
		&MASKING (\uppercase\expandafter{\romannumeral3}) 	 &   72 $\pm$ 67 & $\mathbf{<0.01}$ & 0.713 &   \\
		&NOISE-AWARE (\uppercase\expandafter{\romannumeral4})  &  83 $\pm$ 83 & $\mathbf{<0.01}$ & $\mathbf{<0.01}$ & $\mathbf{<0.01}$ \\
		\hline
		\multirow{4}{*}{1.51} &CNN \cite{gamper2018blind} (\uppercase\expandafter{\romannumeral1}) & 441 $\pm$ 165  &   &   &   \\      
		&RE-NET (\uppercase\expandafter{\romannumeral2}) 	 &   366 $\pm$ 154 & $\mathbf{<0.01}$ &   &   \\
		&MASKING (\uppercase\expandafter{\romannumeral3}) 	 &   265 $\pm$ 105 & $\mathbf{<0.01}$ & $\mathbf{<0.01}$ &   \\
		&NOISE-AWARE (\uppercase\expandafter{\romannumeral4})  &  205 $\pm$ 107 & $\mathbf{<0.01}$ & $\mathbf{<0.01}$ & $\mathbf{<0.01}$ \\
		\hline
	\end{tabular}
	\label{table:ANOVA-Reverb-method}
\end{table}

To investigate the impact of reverberation, the performances of four methods in each room are shown in Figure \ref{figure:boxplot_eachroom_real}. Similar to the simulation results, the estimators tend to overestimate the $T_{60}$ in small reverberation scenarios and underestimate the $T_{60}$ in large reverberation scenarios. And the results in large reverberation scenarios are severely biased.
For small and medium reverberation scenarios (Room 1\#, 2\#, 3\#), RE-NET shows a smaller variance than masking and noise-aware methods. 
A two-way ANOVA is also conducted on the RMSE among different methods and $T_{60}$s. 
Method [$F(3, 11996) = 682.240, p < 0.01$] and $T_{60}$ [$F(2, 11997) = 6336.303, p < 0.01$] have a significant effect on the RMSEs. The interaction between method and $T_{60}$ also shows a significant effect on the RMSEs [$F(6, 11988) = 245.131, p < 0.01$].
To evaluate the performances of methods in different $T_{60}$s, Fisher LSD post hoc test between methods is conducted with different fixed $T_{60}$s and shown in Table \ref{table:ANOVA-Reverb-method}.
In small reverberation scenarios ($T_{60} = 0.32$ s), RE-NET, masking and noise-aware  show no significant difference in between ($p > 0.05$), and they significantly outperform CNN ($p < 0.01$). 
In medium reverberation scenarios ($T_{60} =$ 0.82-0.83 s), RE-NET and masking have no significant difference in between ($p = 0.713$), and they significantly outperform noise-aware and CNN ($p < 0.01$).
In large reverberation scenario ($T_{60} = 1.51$ s), all methods have significant differences in between ($p < 0.01$), and the proposed noise-aware method has a significantly better performance.
The results indicate that $T_{60}$ estimation in large reverberation scenarios are more challenging than in small and medium reverberation scenarios. The noise-aware method significantly improves the estimation accuracy in large reverberation scenarios with the denoising and noise estimation procedures in the first stage, proving the potential of the noise-aware method in challenging (large reverberation and low SNR) scenarios.
}

\section{CONCLUSIONS AND FUTURE WORK} \label{sec:conclu}
Blind $T_{60}$ estimation in noisy and large reverberation enclosures is still a challenging problem. Based on the fact that it is hard to distinguish late reverberation from noise in noisy scenarios, this paper proposes a noise-robust blind reverberation time estimation algorithm based on noise-aware time-frequency masking. 
With the noise-free speech estimation and the noise estimation in the first stage, the deep neural networks can estimate the $T_{60}$ based on the prior speech and noise information in the second stage, which may alleviate some adverse effects of noise.
To evaluate the performance of the proposed method, simulation and real-world acoustic measurements were carried out.
The experimental results show that the proposed noise-aware method outperforms the other baselines especially in low SNR and large reverberation scenarios. The comparison results of training networks with different datasets indicate that training the neural networks with various and complicated noise types greatly improves the robustness of neural networks in unseen noisy scenarios. In terms of the input length of speech, a 3-second duration of input speech is enough for a primary estimation while a longer input length increases the robustness of the algorithm. 
The statistical analysis in real-world testing also indicates that the denoising stage significantly improves the estimation accuracy in noisy scenarios and the noise-aware method shows significantly better performance in challenging scenarios.

In this study, some public available realistic RIRs are collected to augment the training set. However, $T_{60}$s of these RIRs mainly range from 0.3 s to 0.8 s, which results in the degraded performance in large reverberation scenarios. More realistic RIRs with large $T_{60}$s need to be collected or some methods for augmenting the RIRs should be developed in the future work. 

\section{APPENDIX A: ACOUSTIC MEASUREMENT DETAILS}
This section contains a detailed description of the hardware, software and recording method used in the recording procedure which records the reverberant speech and noise in realistic rooms. 

\begin{figure}[t]
	\setlength{\abovecaptionskip}{0.235cm}
	\setlength{\belowcaptionskip}{-0.1cm}
	\centering
	\centerline{\includegraphics[width= 1\columnwidth]{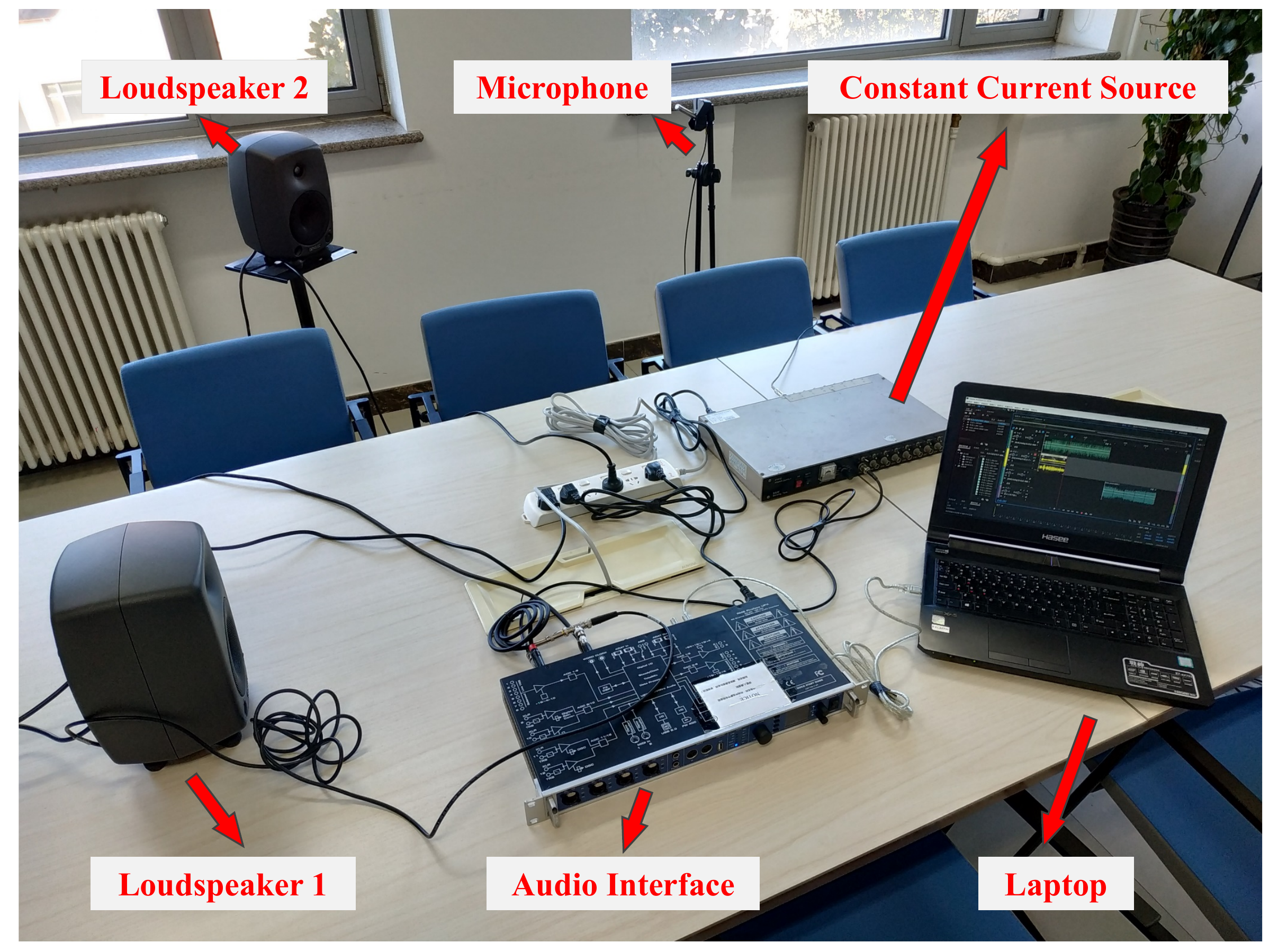}}
	\caption{ Recording session in a meeting room. The recording equipment contains a laptop, an audio interface, a constant current source, a microphone, two loudspeakers and audio lines. }
	\label{figure:experiment_pic}
	\vspace{0cm}
\end{figure}

\subsection{Hardware}

The recording hardware contains a laptop, an audio interface, a constant current source, a microphone, two loudspeakers (one for playing the excitation signal and speech signal, one for playing the noise signal), audio lines and the power sources. The types of hardware used in this experiment are listed as follows:  

Microphone: Prepolarized 1/2'' Type CHZ-213+YG-201 produced by Beijing AcousticSensing Technology Ltd. 

Audio interface: RME Fireface UFX 24 Bit 192 kHz 60 Channel USB/Firewire Audio Interface.

Constant current source: 61012 power supply produced by B\&W Sensing Tech.

laptop: Hasee God of war Notebook Z7-KP7S1.

Loudspeaker: GENELEC 8030B active studio monitor. 

%
Figure \ref{figure:experiment_pic} shows the connected recording equipment in one of the measured rooms.

\subsection{Recording procedure}
{
	The positions of noise source and signal source were fixed and five receiving points were selected as the position of microphone. For each receiving point, the following procedure was performed:
	Five minutes of speech in TIMIT \cite{garofolo1993timit} was used as the clean speech signal. 
	The noise signal contained three noise types from ACE challenge \cite{eaton2016estimation} as well as pink and white noises from NOISEX92 noise dataset \cite{varga1993assessment}. Each noise had a duration of 1 minute.
	For calculating the ground truth of $T_{60}$, two frequently-used excitation signals - Maximum Length Sequence (MLS) \cite{schroeder1979integrated} and Exponential Sine Sweep (ESS) \cite{farina2000simultaneous} techniques were used to measure the RIR.
	The length of the MLS signal was 10.92 s with 48 kHz sampling frequency. The frequency range of the ESS signal was from 20 Hz to 20 kHz with a duration of 20 s.
	RIRs measured with two different techniques were compared carefully and the RIRs using ESS were adopted as they are robust to non-linear distortions \cite{farina2000simultaneous}. 
}

The playback and recording software employed was Audition CC 2019 and the multi-track mode was used. All the signals were recorded using a sample rate of 48 kHz and 32-bit precision. 

\subsection{Test Set Construction}

For the construction of the test set, the recorded reverberant speech was seperated into 4 s utterances with 2 s overlap. For each position in a room, one randomly selected noise from 5 noises was added to each utterance with a SNR selected from 0 dB, 5 dB, 10 dB, 15 dB and 20 dB. 150 noisy reverberant speech signals were used for each position and 5 positions were mesured in each room, yielding 3000 utterances in total. The ground truth of the $T_{60}$ was calculated from the RIR in this position \cite{antsalo2001estimation}.
To validate the calculated $T_{60}$ using RIR , $T_{60}$ was also measured by interrupted noise method \cite{iso20083382} and the mean difference between these two measurements is within 0.02 s.


\section*{Acknowledgements}
This work was supported by National Science Fund of China Under Grant No. 12074403 and No. 11974086.



%
%

\bibliography{enhancefort60}

\begin{thebibliography}{10}
\expandafter\ifx\csname url\endcsname\relax
  \def\url#1{\texttt{#1}}\fi
\expandafter\ifx\csname urlprefix\endcsname\relax\def\urlprefix{URL }\fi
\expandafter\ifx\csname href\endcsname\relax
  \def\href#1#2{#2} \def\path#1{#1}\fi

\bibitem{kuttruff2013room}
H.~Kuttruff, E.~Mommertz, Room acoustics, in: Handbook of engineering
  acoustics, Springer, 2013, pp. 239--267.

\bibitem{iso20083382}
E.~ISO, 3382-2, 2008,“acoustics—measurement of room acoustic
  parameters—part 2: Reverberation time in ordinary rooms,”, International
  Organization for Standardization, Brussels, Belgium (2008).

\bibitem{iso2009acoustics}
I.~3382-1, Acoustics—measurement of room acoustic parameters—part 1:
  Performance spaces, International Standard Organization: Geneva, Switzerland
  (2009).

\bibitem{schroeder1965new}
M.~R. Schroeder, New method of measuring reverberation time, The Journal of the
  Acoustical Society of America 37~(6) (1965) 1187--1188.

\bibitem{ratnam_blind_2003}
R.~Ratnam, D.~L. Jones, B.~C. Wheeler, W.~D. O’Brien~Jr, C.~R. Lansing, A.~S.
  Feng, Blind estimation of reverberation time, The Journal of the Acoustical
  Society of America 114~(5) (2003) 2877--2892, publisher: Acoustical Society
  of America.

\bibitem{lollmann_estimation_2008}
H.~W. Löllmann, P.~Vary, Estimation of the reverberation time in noisy
  environments, in: Proc. of {Intl}. {Workshop} on {Acoustic} {Echo} and
  {Noise} {Control} ({IWAENC}), Citeseer, 2008.

\bibitem{eaton_noise-robust_2013}
J.~Eaton, N.~D. Gaubitch, P.~A. Naylor, Noise-robust reverberation time
  estimation using spectral decay distributions with reduced computational
  cost, in: 2013 {IEEE} {International} {Conference} on {Acoustics}, {Speech}
  and {Signal} {Processing}, IEEE, 2013, pp. 161--165.

\bibitem{prego_blind_2015}
T.~d.~M. Prego, A.~A. de~Lima, R.~Zambrano-López, S.~L. Netto, Blind
  estimators for reverberation time and direct-to-reverberant energy ratio
  using subband speech decomposition, in: 2015 {IEEE} workshop on applications
  of signal processing to audio and acoustics ({WASPAA}), IEEE, 2015, pp. 1--5.

\bibitem{loellmann_single-channel_2015}
H.~Loellmann, A.~Brendel, P.~Vary, W.~Kellermann, Single-channel
  maximum-likelihood {T60} estimation exploiting subband information, arXiv
  preprint arXiv:1511.04063 (2015).

\bibitem{cox_extracting_2001}
T.~J. Cox, F.~Li, P.~Darlington, Extracting room reverberation time from speech
  using artificial neural networks, Journal of the audio engineering society
  49~(4) (2001) 219--230, publisher: Audio Engineering Society.

\bibitem{parada_evaluating_2015}
P.~P. Parada, D.~Sharma, T.~van Waterschoot, P.~A. Naylor, Evaluating the
  non-intrusive room acoustics algorithm with the {ACE} challenge, arXiv
  preprint arXiv:1510.04616 (2015).

\bibitem{xiong_joint_2015}
F.~Xiong, S.~Goetze, B.~T. Meyer, Joint estimation of reverberation time and
  direct-to-reverberation ratio from speech using auditory-inspired features,
  arXiv preprint arXiv:1510.04620 (2015).

\bibitem{eaton2016estimation}
J.~Eaton, N.~D. Gaubitch, A.~H. Moore, P.~A. Naylor, Estimation of room
  acoustic parameters: The ace challenge, IEEE/ACM Transactions on Audio,
  Speech, and Language Processing 24~(10) (2016) 1681--1693.

\bibitem{lee2016blind}
M.~Lee, J.-H. Chang, Blind estimation of reverberation time using deep neural
  network, in: 2016 IEEE International Conference on Network Infrastructure and
  Digital Content (IC-NIDC), IEEE, 2016, pp. 308--311.

\bibitem{deng2020online}
S.~Deng, W.~Mack, E.~A. Habets, Online blind reverberation time estimation
  using crnns, Proc. Interspeech 2020 (2020) 5061--5065.

\bibitem{srivastava2021blind}
P.~Srivastava, A.~Deleforge, E.~Vincent, Blind room parameter estimation using
  multiple-multichannel speech recordings, arXiv preprint arXiv:2107.13832
  (2021).

\bibitem{gaubitch2012performance}
N.~D. Gaubitch, H.~W. Loellmann, M.~Jeub, T.~H. Falk, P.~A. Naylor, P.~Vary,
  M.~Brookes, Performance comparison of algorithms for blind reverberation time
  estimation from speech, in: IWAENC 2012; International Workshop on Acoustic
  Signal Enhancement, VDE, 2012, pp. 1--4.

\bibitem{prego2015blind}
T.~d.~M. Prego, A.~A. de~Lima, R.~Zambrano-L{\'o}pez, S.~L. Netto, Blind
  estimators for reverberation time and direct-to-reverberant energy ratio
  using subband speech decomposition, in: 2015 IEEE workshop on applications of
  signal processing to audio and acoustics (WASPAA), IEEE, 2015, pp. 1--5.

\bibitem{reddy2021icassp}
C.~K. Reddy, H.~Dubey, V.~Gopal, R.~Cutler, S.~Braun, H.~Gamper, R.~Aichner,
  S.~Srinivasan, Icassp 2021 deep noise suppression challenge, in: ICASSP
  2021-2021 IEEE International Conference on Acoustics, Speech and Signal
  Processing (ICASSP), IEEE, 2021, pp. 6623--6627.

\bibitem{li2021icassp}
A.~Li, W.~Liu, X.~Luo, C.~Zheng, X.~Li, Icassp 2021 deep noise suppression
  challenge: Decoupling magnitude and phase optimization with a two-stage deep
  network, in: ICASSP 2021-2021 IEEE International Conference on Acoustics,
  Speech and Signal Processing (ICASSP), IEEE, 2021, pp. 6628--6632.

\bibitem{liu2021know}
W.~Liu, A.~Li, Y.~Ke, X.~Zheng, Chengshi amd~Li, Know your enemy, know yourself
  a unified two-stage framework for speech enhancement, Proc. Interspeech 2021
  (2021).

\bibitem{zhang2019robust}
W.~Zhang, Y.~Zhou, Y.~Qian, Robust doa estimation based on convolutional neural
  network and time-frequency masking., in: INTERSPEECH, 2019, pp. 2703--2707.

\bibitem{wang2018robust}
Z.-Q. Wang, X.~Zhang, D.~Wang, Robust speaker localization guided by deep
  learning-based time-frequency masking, IEEE/ACM Transactions on Audio,
  Speech, and Language Processing 27~(1) (2018) 178--188.

\bibitem{gamper2018blind}
H.~Gamper, I.~J. Tashev, Blind reverberation time estimation using a
  convolutional neural network, in: 2018 16th International Workshop on
  Acoustic Signal Enhancement (IWAENC), IEEE, 2018, pp. 136--140.

\bibitem{ronneberger2015u}
O.~Ronneberger, P.~Fischer, T.~Brox, U-net: Convolutional networks for
  biomedical image segmentation, in: International Conference on Medical image
  computing and computer-assisted intervention, Springer, 2015, pp. 234--241.

\bibitem{dauphin2017language}
Y.~N. Dauphin, A.~Fan, M.~Auli, D.~Grangier, Language modeling with gated
  convolutional networks, in: International conference on machine learning,
  PMLR, 2017, pp. 933--941.

\bibitem{tan2019learning}
K.~Tan, D.~Wang, Learning complex spectral mapping with gated convolutional
  recurrent networks for monaural speech enhancement, IEEE/ACM Transactions on
  Audio, Speech, and Language Processing 28 (2019) 380--390.

\bibitem{li2021two}
A.~Li, W.~Liu, C.~Zheng, C.~Fan, X.~Li, Two heads are better than one: A
  two-stage complex spectral mapping approach for monaural speech enhancement,
  IEEE/ACM Transactions on Audio, Speech, and Language Processing 29 (2021)
  1829--1843.

\bibitem{ruder2017overview}
S.~Ruder, An overview of multi-task learning in deep neural networks, arXiv
  preprint arXiv:1706.05098 (2017).

\bibitem{wang2017joint}
Q.~Wang, J.~Du, L.-R. Dai, C.-H. Lee, Joint noise and mask aware training for
  dnn-based speech enhancement with sub-band features, in: 2017 Hands-free
  Speech Communications and Microphone Arrays (HSCMA), IEEE, 2017, pp.
  101--105.

\bibitem{ulyanov2016instance}
D.~Ulyanov, A.~Vedaldi, V.~Lempitsky, Instance normalization: The missing
  ingredient for fast stylization, arXiv preprint arXiv:1607.08022 (2016).

\bibitem{he2015delving}
K.~He, X.~Zhang, S.~Ren, J.~Sun, Delving deep into rectifiers: Surpassing
  human-level performance on imagenet classification, in: Proceedings of the
  IEEE international conference on computer vision, 2015, pp. 1026--1034.

\bibitem{luo2019conv}
Y.~Luo, N.~Mesgarani, Conv-tasnet: Surpassing ideal time--frequency magnitude
  masking for speech separation, IEEE/ACM transactions on audio, speech, and
  language processing 27~(8) (2019) 1256--1266.

\bibitem{li2020speech}
A.~Li, M.~Yuan, C.~Zheng, X.~Li, Speech enhancement using progressive
  learning-based convolutional recurrent neural network, Applied Acoustics 166
  (2020) 107347.

\bibitem{yuan2015speech}
W.~Yuan, B.~Xia, A speech enhancement approach based on noise classification,
  Applied Acoustics 96 (2015) 11--19.

\bibitem{paul1992design}
D.~B. Paul, J.~Baker, The design for the wall street journal-based csr corpus,
  in: Speech and Natural Language: Proceedings of a Workshop Held at Harriman,
  New York, February 23-26, 1992, 1992.

\bibitem{habets2006room}
E.~A. Habets, Room impulse response generator, Technische Universiteit
  Eindhoven, Tech. Rep 2~(2.4) (2006) 1.

\bibitem{antsalo2001estimation}
P.~Antsalo, A.~Makivirta, V.~Valimaki, T.~Peltonen, M.~Karjalainen, Estimation
  of modal decay parameters from noisy response measurements, in: Audio
  Engineering Society Convention 110, Audio Engineering Society, 2001.

\bibitem{murphy2010openair}
D.~T. Murphy, S.~Shelley, Openair: An interactive auralization web resource and
  database, in: Audio Engineering Society Convention 129, Audio Engineering
  Society, 2010.

\bibitem{kinoshita2013reverb}
K.~Kinoshita, M.~Delcroix, T.~Yoshioka, T.~Nakatani, E.~Habets, R.~Haeb-Umbach,
  V.~Leutnant, A.~Sehr, W.~Kellermann, R.~Maas, et~al., The reverb challenge: A
  common evaluation framework for dereverberation and recognition of
  reverberant speech, in: 2013 IEEE Workshop on Applications of Signal
  Processing to Audio and Acoustics, IEEE, 2013, pp. 1--4.

\bibitem{nakamura2000acoustical}
S.~Nakamura, K.~Hiyane, F.~Asano, T.~Nishiura, T.~Yamada, Acoustical sound
  database in real environments for sound scene understanding and hands-free
  speech recognition (2000).

\bibitem{reddy2020interspeech}
C.~K. Reddy, E.~Beyrami, H.~Dubey, V.~Gopal, R.~Cheng, R.~Cutler,
  S.~Matusevych, R.~Aichner, A.~Aazami, S.~Braun, et~al., The interspeech 2020
  deep noise suppression challenge: Datasets, subjective speech quality and
  testing framework, arXiv preprint arXiv:2001.08662 (2020).

\bibitem{garofolo1993timit}
J.~S. Garofolo, Timit acoustic phonetic continuous speech corpus, Linguistic
  Data Consortium, 1993 (1993).

\bibitem{varga1993assessment}
A.~Varga, H.~J. Steeneken, Assessment for automatic speech recognition: Ii.
  noisex-92: A database and an experiment to study the effect of additive noise
  on speech recognition systems, Speech communication 12~(3) (1993) 247--251.

\bibitem{kingma2014adam}
D.~P. Kingma, J.~Ba, Adam: A method for stochastic optimization, arXiv preprint
  arXiv:1412.6980 (2014).

\bibitem{schroeder1979integrated}
M.~R. Schroeder, Integrated-impulse method measuring sound decay without using
  impulses, The Journal of the Acoustical Society of America 66~(2) (1979)
  497--500.

\bibitem{farina2000simultaneous}
A.~Farina, Simultaneous measurement of impulse response and distortion with a
  swept-sine technique, in: Audio Engineering Society Convention 108, Audio
  Engineering Society, 2000.

\end{thebibliography}

\end{document}